# Magnetic Flux Transport by turbulent reconnection in astrophysical flows

E. M. de Gouveia Dal Pino[1], M. R. M. Leão[1], R. Santos-Lima[1],

G. Guerrero[2], G. Kowal[1], and A. Lazarian[3]

[1] Instituto de Astronomia, Geofísica e Ciências Atmosféricas, Universidade de São Paulo, R. do Matão, 1226, São Paulo, SP 05508-090, Brazil

[2] Hansen Experimental Physics Laboratory, Stanford University, Stanford, CA 94305, USA

[3] Department of Astronomy, University of Wisconsin, Madison,WI 53706, USA

Email: dalpino@astro.iag.usp.br

**Abstract**  The role of MHD turbulence in astrophysical environments is still highly debated. An important question that permeates this debate is the transport of magnetic flux. This is particularly important, for instance, in the context of star formation. When clouds collapse gravitationally to form stars, there must be some magnetic flux transport. otherwise the new born stars would have magnetic fields several orders of magnitude larger than the observed ones. Also, the magnetic flux that is dragged in the late stages of the formation of a star can remove all the rotational support from the accretion disk that grows around the protostar. The efficiency of the mechanism which is often invoked to allow the transport of magnetic fields in the different stages of star formation, namely, the ambipolar diffusion, has been lately put in check. We here discuss an alternative mechanism for magnetic flux transport which is based on turbulent fast magnetic reconnection. We review recent results obtained from 3D MHD numerical simulations that indicate that this mechanism is very efficient for decoupling and transport magnetic flux from the inner denser regions to the outskirts of collapsing clouds in the different stages of star formation. We also discuss this mechanism in the context of dynamo processes and speculate that it can play a role both in the solar dynamo and in accretion disk dynamo processes.

## 1. Introduction

Astrophysical flows are known to be turbulent and magnetized. The specific role played by MHD turbulence in different astrophysical flows is still highly debated. One question that frequently permeates these debates is the diffusion of the magnetic field. The conductivity is high enough to make the Ohmic diffusion negligible in the scales typically involved in most of the astrophysical fluids. This means that the ideal MHD (or "frozen-in") approximation is appropriate for most of these environments. However, without considering diffusive mechanisms that can violate the magnetic flux freezing, one faces several problems. For instance, simple estimates show that if all the magnetic flux were brought together with the material that collapsed to form a star in a molecular cloud, then the magnetic fields in protostars should be several orders of magnitude larger than the observed ones. This is often referred as the "magnetic flux problem" in star formation. Likewise, magnetic flux transport is known to be a necessary ingredient in dynamo processes.

We here discuss a *new* mechanism for magnetic flux transport which is based on turbulent fast magnetic reconnection. In section 2, we summarize the theoretical grounds of this transport mechanism. In section 3, we review recent results of a study of this mechanism applied to star formation. Then, in section 4, we speculate that this mechanism can also play a role in dynamo processes, both in accretion disks and the solar dynamo, and in section 5 we draw our conclusions.

**2. Turbulent magnetic reconnection: theoretical grounds**

The magnetic diffusion mechanism that we will address is a process deeply rooted in the microphysics behavior of the magnetic fields in highly conductive flows. Textbooks characterize these flows by the Lundquist number $S = L_X V_A /\eta$, where $\eta$ is the Ohmic diffusivity, $L_X$ is a typical scale of the system and $V_A$ is the Alfvén velocity, For astrophysical systems $L_X$ is in general very large and therefore, $S \gg 1$ which makes magnetic diffusion negligible. However, one may ask: does magnetic field remain absolutely frozen-in within highly ionized astrophysical fluids? The answer to this question relies on magnetic reconnection.

Magnetic reconnection occurs when two magnetic fluxes of opposite polarity encounter each other. In the presence of finite magnetic resistivity, the converging magnetic lines annihilate at the discontinuity surface and a current sheet forms there. In the standard Sweet-Parker (S-P) model the velocity at which the two converging fluxes reconnect is given by $v_{rec} \approx v_A S^{-1/2}$, where in this case $L_X$ gives the length of the reconnection layer (see Figure 1). Because S is large for Ohmic resistivity values (e.g., for the ISM, $S \sim 10^{16}$), the Sweet-Parker reconnection is very *slow*. In other words, since all the matter moving with the speed $V_{rec}$ over the scale $L_X$ must be ejected with the Alfvén velocity through a thin slot ($\Delta$), the disparity between the typical scales $L_X$ and the outflow thickness $\Delta$, which in turn is determined by microphysics, i.e. the resistivity, makes the Sweet-Parker reconnection rate negligibly small.

However, observations indicate that magnetic reconnection must be *fast* in some circumstances (e.g., solar flares). Lazarian & Vishniac (1999) proposed a model for fast reconnection that is independent of the resistivity. The model appeals to the ubiquitous astrophysical turbulence as a universal trigger of fast reconnection. When turbulence is present within the current sheet, the outflow region $\Delta$ gets determined by magnetic field wandering and therefore, becomes independent of the resistivity (see Figure 1, middle panel). It allows the formation of a thick volume filled with several reconnected small magnetic fluctuations which make the reconnection fast. This model was successfully tested numerically by Kowal et al. (2009) (see Figure 1, bottom panel). This challenges the well-rooted concept of magnetic field frozenness for the case of turbulent fluids and provides an interesting way of removing magnetic flux out of astrophysical flows, e.g. star formation regions (Lazarian 2005; Santos-Lima et al. 2010; 2012; de Gouveia Dal Pino et al. 2011; Lazarian 2011), accretion disks, or the solar dynamo.

It must be remarked that numerical effects are always a concern when dealing with numerical simulations involving reconnection and magnetic field diffusion. However, the high resolution numerical tests of magnetic reconnection performed by Kowal et al. (2009), showed that in the presence of turbulence the local non-linear enhancements of resistivity are not important. This is a confirmation that the turbulent reconnection diffusion that we observe in our simulations (see below and also Santos-Lima et al. 2010; 2012) is a real effect and not a numerical artefact. Analytical studies summarized in Eyink et al. (2011) also support the notion that magnetic fields are generically not frozen-in when conductive fluids are turbulent. In view of these studies, we can conclude that the concept of reconnection diffusion looks very natural and ubiquitous.

In the following sections, we will review recent numerical studies of the application of this diffusion mechanism to star and protostellar disk formation, and also discuss its application to dynamo processes in accretion disks and the sun.

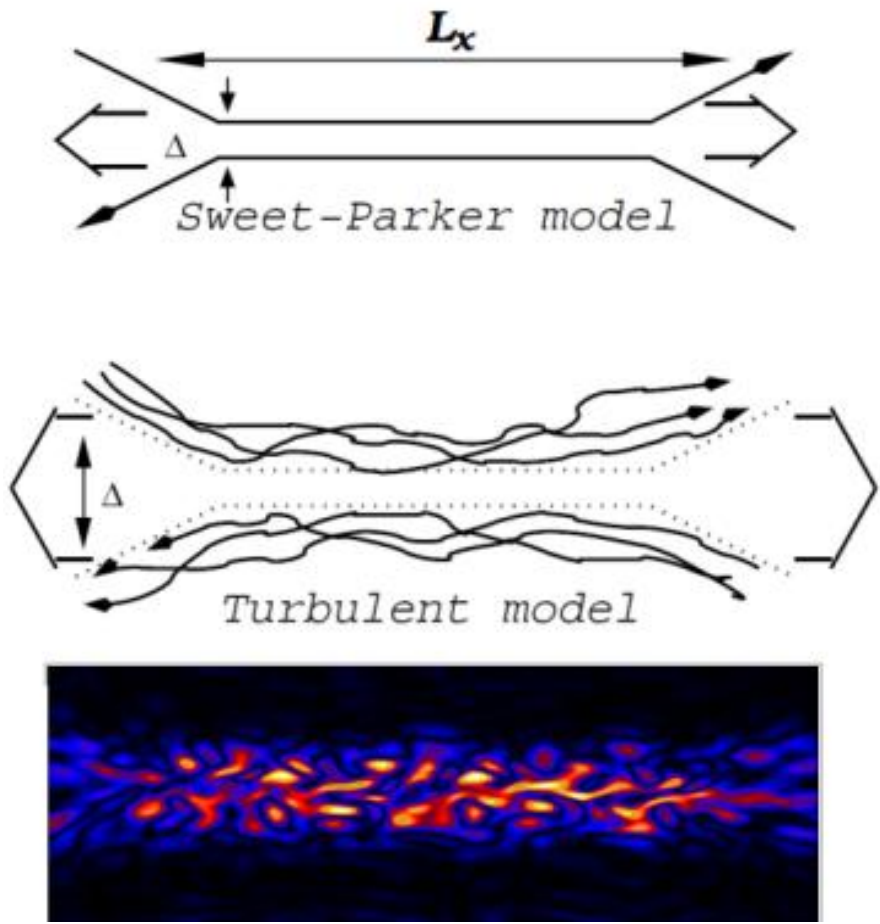

Figure 1.— Top panel: Sweet–Parker model of reconnection. The outflow is limited by a thin slot $\Delta$ determined by Ohmic diffusivity. The other scale is an astrophysical scale $L_x \gg \Delta$. Middle panel: fast reconnection model in presence of turbulence according to Lazarian & Vishiniac (1999) (extracted from Lazarian et al. 2004). Bottom panel: 3D MHD numerical simulation of fast turbulent reconnection (from Kowal et al. 2009).

**3. The magnetic flux transport problem is star formation**

To address the magnetic flux transport problem in the framework of star formation, researchers usually invoke the ambipolar diffusion (AD) mechanism (e.g. Mestel & Spitzer 1956; Mouschovias 1979; Li et al. 2008). In principle, AD allows magnetic flux to be redistributed during the collapse in low ionization regions as the result of the differential motion between the ionized and the neutral gas. Recent advances in the theory, however, have been putting in check the efficiency of this diffusion process in real systems. Shu et al. (2006), for instance, explored the accretion phase in low-mass star formation and concluded that there should exist an effective diffusivity about three orders of magnitude larger than the classic Ohmic diffusivity in order to allow an efficient magnetic flux transport to occur. They found that AD could work in principle, but only under special circumstances, considering specific dust grain sizes. In other words, there is still no consensus if AD alone is high enough to solve the magnetic flux transport problem in collapsing flows. (see also alternative views in Li et al. 2011).

*3.1 Magnetic flux transport by turbulent reconnection in the early stages of star formation*

We have recently explored the role of turbulent reconnection on the transport of magnetic field flux from the central, denser regions of a molecular cloud to outside, in order to follow the cloud gravitational collapse (Santos-Lima et al. 2010).

Molecular clouds are known to be turbulent and magnetized They have a variety of structures in all scales and turbulence rules their structuring and fragmentation and probably plays an important role during most the different stages star formation (Mac Low & Klessen 2004); McKee & Ostriker 2007; Mac Low & Klessen 2004; Vazquez-Semadeni et al. 2005). A dominant source of turbulence injection is possibly supernova shocks (see Melioli et al. 2006; Leão et al. 2009 and references therein).

We performed 3-D MHD simulations of ISM clouds considering a central gravitational field provided by embedded stars and introducing forced turbulence. We employed a shock-capturing Godunov-type code with an HLL solver to integrate the fluxes and a second order Runge-Kutta to integrate in time (Kowal et al. 2007; Santos-Lima et al. 2010, 2012). We considered an isotropic, non-helical, solenoidal, delta correlated in time turbulent forcing. This forcing acts in a thin shell around the wave number $k = 2.5(2\pi/L)$, so that the scale of turbulence injection $l_{inj}$ is about 2.5 times smaller than the computational domain size L. In all examples below, transonic, sub-Alfvénic turbulence with an rms velocity around unity was injected in the system which was then let to evolve.

When compared with MHD simulations without turbulence, those with turbulence revealed a decrease of the magnetic flux-to-mass ratio as density at the center of the gravitational potential increases. The magnetic flux is transported to the outskirts of the cloud by turbulent reconnection. We observed this effect both when starting with initial equilibrium distributions of gas and magnetic field and when following the evolution of dynamically unstable configurations. Thus the process of turbulent magnetic field removal is applicable both to quasi-static subcritical molecular clouds and collapsing supercritical ones. The increase of the gravitational potential, as well as the decrease of magnetization of the gas showed an increase of the decoupling between the mass and the magnetic flux in the saturated final state of the simulations, supporting the notion that turbulent diffusivity relaxes the magnetic field+gas system in the gravitational field to its minimal energy state (Santos-Lima et al. 2010).

More recently, we have been exploring more realistic systems including the effects of the gas self-gravity in the clouds. The stability of a cloud supported by the magnetic pressure may be quantified by the mass-to-magnetic flux ratio, $M/\Phi \sim N/B$, where M is the cloud mass, $\Phi$ is the magnetic flux, B is the magnetic field, and N is the column density. This ratio defines to what extent a static magnetic field can support a cloud against gravitational collapse (e.g., Nakano & Nakamura 1978; Crutcher 2008). When this ratio exceeds a critical value above which gravity overpasses the magnetic and turbulent forces, then the cloud or cloud core is able to collapse.

We considered an initially spherical cloud clump with a central gravitational potential that mimics a small group of embedded stars. It was put in the middle of a homogeneous magnetized background, then a violent initial contraction of the gas takes place for a short period of time (of the order of the free-fall time), after which due to the presence of the magnetic field and the turbulence injected, the system evolves more smoothly. The system is simulated inside a cubic domain with periodic boundaries. The existence of several clumps in the interior of a typical giant molecular cloud allows us to use periodic boundaries for our setup. For the sake of simplicity, we employ an isothermal equation of state, with a single temperature for the whole system (see more details in Leao et al. 2012, in prep.).

Figure 2 shows examples of this sort of simulations considering conditions appropriate to molecular clouds. The left panels show logarithmic density maps of the central slices of the simulated models after 100 Myrs. The middle column panels compare the time evolution of the magnetic field-to-density ratio, normalized by the average value of this ratio, inside a central sphere with radius r = 0.3pc which represents the core of the cloud, both for the models with turbulence (red-dashed line) and the ones without turbulence (black-solid line). After an initial rapid decrease caused by the relaxation of the system, the magnetic field-to-density ratio remains nearly constant in the laminar cases (i.e., without turbulence)[1], while in the cases with turbulence there is a clear decrease of it. This result indicates that there was magnetic flux transport from the denser, more massive central regions to the less dense regions outside of the cloud cores. This effect is particularly more pronounced in the bottom panel model which has larger initial gas density than the other models and therefore, is under the influence of larger self-gravity. Comparing the top and middle models which have the same initial gas density, the one with larger stellar potential (top model) shows a larger decoupling between the magnetic flux and the mass density. All the models above are initially subcritical clouds, that is, they have an initial mass-to-flux ratio which is smaller than the critical value necessary for the cloud to collapse gravitationally. We have also computed the time evolution of this ratio for the three turbulent models and found that the one with larger initial density (bottom model) is the only one that becomes supercritical and therefore, able to continue the collapse to form stars, while the one with a larger stellar gravitational potential (top model) has only approached the critical value.

The results above suggest that self-gravity has an important effect as it facilitates the gas infall and therefore, the decoupling of the magnetic field that is more easily removed to the outer regions of the collapsing cloud core. The results also suggest that an increase in self-gravity is more important than an increase in the stellar gravitational potential in order to produce magnetic flux transport by turbulent reconnection. However, we found that turbulence is able to remove magnetic flux from collapsing, self-gravitating clouds and make them supercritical within a narrow range of densities (for clouds with ~50 solar mass, 10 < n < 100

---

[1] We note that the oscillations observed in these plots (which are slightly stronger in the laminar models) are acoustic oscillations of the cloud due to the fact that the virialization time of these systems is larger than the simulated period.

cm$^{-3}$; see Leão et al. 2012, for more details). At the same time, this result is compatible with the known low efficiency of star formation in the Galaxy

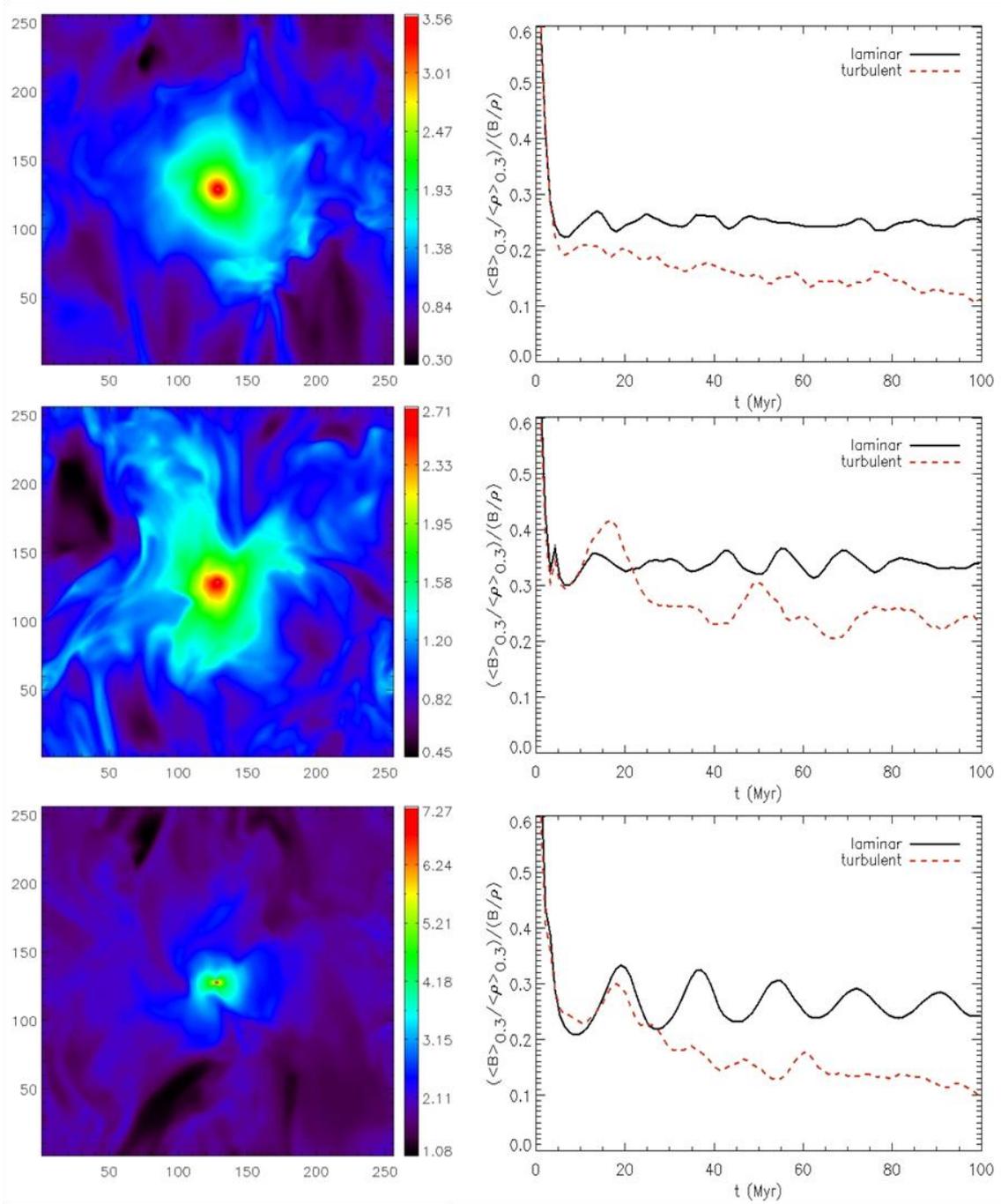

Figure 2. Left panels show density maps of the central slices of the cloud evolution for turbulent models at t = 100 Myrs. Right panels show the temporal evolution of the magnetic field-to-density ratio at the cloud core region (within a radius = 0.3pc) normalized by the average value in the system, both for turbulent (red-dashed lines) and laminar (black continuous lines) models. Top: model has a stellar potential $M_{pot}$ = 61.1 solar mass. Center and bottom models have $M_{pot}$ = 40.7 solar mass. Top and middle models have initial densities n=10 cm$^{-3}$, and the bottom model has n=90 cm$^{-3}$. All models have a thermal to magnetic pressure ratio β = 3.0 (see also Leão et al. 2012).

*3.1 Magnetic flux transport by turbulent reconnection in the late stages of star formation*

The late stages of star formation are not fully understood either (see Krasnopolsky et al. 2011 for a recent review). Former studies have shown that the observed embedded magnetic field in molecular cloud cores (Troland & Crutcher 2008) is high enough to inhibit the formation of rationally supported disks during the main protostellar accretion phase of low mass stars, if ideal MHD applies. This has been known as the magnetic *braking* problem (see e.g., Galli et al. 2006; Price & Bate 2007; Hennebelle & Fromang 2008; Mellon & Li 2008). For realistic levels of core magnetization and ionization, recent work has shown that again AD does not seem to be sufficient to weaken the magnetic braking in order to allow rotationally supported disks to form. In some cases, the magnetic braking has been found to be even enhanced by AD (Mellon & Li 2009; Krasnopolsky & Konigl 2002; Basu & Mouschovias 1995; Hosking & Whitworth 2004; Duffin & Pudritz 2009; Li et al. 2011). These findings motivated Krasnopolsky et al. (2010) (see also Li et al. 2011) to examine whether Ohmic dissipation could be effective in weakening the magnetic braking. They claimed that in order to enable the formation of a rotationally supported disks during the protostellar mass accretion phase an enhanced resistivity a few orders of magnitude higher than the classic Ohmic resistivity would be required (Krasnopolsky et al. 2010). On the other hand, Machida et al. (2010) (see also Inutsuka et al. 2010; Machida et al. 2011) performed core collapse three-dimensional simulations and found that with just the Ohmic resistivity, a massive, rotationally supported disk can form but the process is slow, and one has to wait for over $10^5$ yr for this to occur.

While this question on the effectiveness of the Ohmic diffusion in the disk formation still deserves further testing, considering the success of the turbulent magnetic reconnection discussed before to remove magnetic flux in the early stages of star formation, we have also investigated this mechanism during the late phases of the protostellar disk formation. We showed by means of 3D MHD simulations that the diffusivity arising from turbulent magnetic reconnection is able to transport magnetic flux to the outskirts of the (cloud core) disk progenitor at time scales compatible with the collapse. In just a few $10^4$ yr, a rotationally supported disk forms around the protostar of dimensions ~100 AU, with a nearly Keplerian profile as required by the observations. Since MHD turbulence is expected to be present in protostellar disks, this is a natural mechanism for removing magnetic flux excess and allowing the formation of these disks (Santos-Lima et al. 2012; de Gouveia Dal Pino et al. 2011).

**4. Investigation of the role of turbulent reconnection magnetic flux transport in dynamo mechanisms**

In this section, we discuss the potential role of turbulent reconnection mechanism in the framework of dynamo processes, both in accretion disks and in the sun, on very preliminary basis. In forthcoming work, we will explore quantitatively the predictions here made.

*4.1. In the framework of the solar dynamo*

It is well known that the turbulent motions in the convective layer are a key ingredient in the solar dynamo. Therefore, besides the inductive role that helical turbulence might play in amplifying the solar magnetic field, it is natural to expect that the mechanism of turbulent reconnection transport of magnetic flux discussed above can also have an important role in the solar dynamo cycle.

Over many years it was thought that the sunspots migration pattern observed at the solar surface would correspond to the propagation of a dynamo wave. This requires negative kinetic helicity in the northern hemisphere (as observed) and a radial differential rotation profile increasing towards the center of the Sun. However, helioseismology found that in the bulk of the convection zone the radial profile of the differential rotation is almost flat, but rapidly decreases in the inner interface (named tachocline) between the convective and the radiative layers. The migration pattern of a dynamo wave with such characteristics fails to reproduce the observations. Today it is known that near the solar surface there is a thin layer where there is a negative velocity shear (i.e., the velocity increases inwards). This layer alone could give the proper surface shape to a dynamo wave generated by turbulence in the entire convection zone (Brandenburg, 2005). Another class of dynamos assumes that the magnetic field observed in the sunspots is formed at the tachocline. Since the shear there at lower latitudes is negative, these models rely on the meridional circulation (a large scale flow occurring in the r and $\theta$ directions) to explain the observed sunspots migration. Nevertheless, observations indicate that this flow is rather incoherent, changing from one solar cycle to the next and even during the same cycle. Global numerical simulations have not succeeded either in obtaining a well defined meridional circulation flow. In Guerrero & de Gouveia Dal Pino (2008), it was proposed that turbulent pumping could be the mechanism responsible for the transport of the magnetic flux in the observed directions. Turbulent pumping is an advective transport coefficient of the electromotive force in the mean-field MHD model. The role that turbulent magnetic reconnection can play in this process still remains to be understood.

In the upper layers near the solar surface, it might be examined if turbulent reconnection combined with latitudinal (and radial) shear motions can help the deposition of magnetic flux near the equator, as observed. Its contribution in the deeper convective layers where the turbulent convection is mostly anisotropic must also be tested, particularly in order to understand its behavior towards the pumping and storage of the magnetic field in the tachocline.

The suppression of the turbulence and so of the magnetic diffusivity (as well as other turbulent processes) with the increase of the magnetic field (the so called $\eta$-quenching) is another interesting issue to be explored in depth (e.g., Rüdiger, Kitchatinov, Küker & Schultz 1994; Tobias 1996; Guerrero, Dikipati & de Gouveia Dal Pino 2009).

In forthcoming work, we will investigate the potential role of the turbulent reconnection transport mechanism in the solar dynamo, in its linear and non-linear phases, by means of 3D MHD simulations using the same numerical tools and codes as described in the previous section, including all the essential ingredients, such as stratification and rotation, besides the injection of forced turbulence to simulate the convective layer.

*4.2 In the framework of accretion disks*

The application of the reconnection diffusion concept to protostellar disk formation (section 3.2) and, in a more general framework, to accretion disks, is natural as the disks are expected to be turbulent. In fact, the well investigated magneto-rotational instability (MRI; Chandrasekhar 1960; Balbus & Hawley 1991) is effective not only in the transport of angular momentum in the disk, but also in triggering turbulence. This in turn, may help in the amplification of the magnetic fields in the disk in a dynamo process (Livio, Pringle & King 2003). However, once a large scale magnetic field is established it may be strong enough to inhibit the MRI which then stops to operate.

On the other hand, as discussed before in the process of disk formation, if turbulence is still present, this large scale magnetic field can be efficiently removed from the inner, denser regions of the accretion disk to the outer regions by the action of turbulent reconnection transport and then the MRI instability can be resumed, initiating a new phase of the dynamo process. This interplay between the MRI and the turbulent reconnection transport of magnetic flux in accretion disks in a dynamo process by means of fully 3D MHD simulations will be explored elsewhere. Nonetheless, it should be noted that former studies of the injection of turbulence in accretion disks have shown that turbulence may be ineffective to magnetic flux diffusion outward (Rothstein & Lovelace 2008).

## 5. Summary and conclusions

We have reviewed recent results regarding the transport of magnetic flux in astrophysical conducting flows in the presence of turbulence. We discussed a new transport mechanism which is based on the fact that, in the presence of turbulence, magnetic reconnection becomes fast and therefore, very effective to diffuse the magnetic flux (Lazarian & Vishniac 1999; Kowal et al. 2009).

We investigated this turbulent reconnection mechanism in the context of star formation, by means of high resolution 3D MHD numerical simulations, from the early stages of molecular cloud gravitational collapse to the late stages when a Keplerian accretion disk grows around the protostar. We found that this mechanism is very efficient to transport the magnetic flux excess from the inner denser regions to the outskirts of the collapsing system in time scales compatible with the gravitational collapse (Santos-Lima et al. 2012; de Gouveia Dal Pino et al. 2011; Leão et al. 2012). Since turbulence is present in these systems, this mechanism provides a natural way to transport magnetic flux. Besides, it dismisses the necessity of postulating an artificial hypothetical increase of the Ohmic resistivity, as discussed in the literature, and calls for reconsidering the relative role of ambipolar diffusion in the processes of star and planet formation

Finally, we have argued that turbulent fast reconnection can also play a role during the development of dynamo processes both in the more general framework of accretion disks and in the solar dynamo. In forthcoming work we intend to examine in depth these ideas, particularly be means of fully 3D MHD numerical studies.

*Acknowledgements* This work was partially supported by Brazilian grants from FAPESP (2006/50654-3 and 2007/04551-0) and CNPq (140110/2008-9 and 306598/2009-4), and also by funding from the NORDITA program "Dynamo, Dynamical Systems and Topology". EMGDP also acknowledges Axel Brandenbourg and Alexander Kosovichev for their kind hospitality during her stay in NORDITA in July 2011, where part of the ideas of this work were developed.